\documentclass{aa}

\usepackage{graphicx,times}
\def\simlt {\lower.5ex\hbox{$\; \buildrel < \over \sim \;$}} 
\def\simgt{\lower.5ex\hbox{$\; \buildrel > \over \sim \;$}}

\begin{document}
\thesaurus{11         
              (11.09.1 NGC~5548; 
               11.19.1           
               11.17.1           
               13.25.2)}         

  \title{X-ray absorption lines in the Seyfert 1 galaxy NGC~5548 discovered
         with Chandra-LETGS}

\author{ J.S. Kaastra \inst{1}
         \and
         R. Mewe \inst{1}
         \and
         D.A. Liedahl \inst{2}
         \and
         S. Komossa \inst{3}
         \and
         A.C. Brinkman \inst{1}
         }
  
\offprints{J.S. Kaastra}
\mail{J.Kaastra@sron.nl}

\institute{ SRON Laboratory for Space Research
              Sorbonnelaan 2, 3584 CA Utrecht, The Nether\-lands 
              \and
              Physics Department,
              Lawrence Livermore National Laboratory, 
              P.O. Box 808, L-11, Livermore, CA 94550, USA
              \and
              Max Planck Institut f\"ur Extraterrestrische Physik,
              Postfach 1603, D-85740 Garching, Germany}

\date{Received 1-Feb-2000 / Accepted 4-Feb-2000 }

\maketitle

\begin{abstract}

We present for the first time a high-resolution X-ray spectrum of a Seyfert
galaxy.  The Chandra-LETGS spectrum of NGC~5548 shows strong, narrow absorption
lines from highly ionised species (the H-like and He-like ions of C, N, O, Ne,
Na, Mg, Si, as well as \ion{Fe}{xiv} -- \ion{Fe}{xxi}).  The lines are
blueshifted by a few hundred km/s.  The corresponding continuum absorption edges
are weak or absent.  The absorbing medium can be modelled by an outflowing, thin
and warm shell in photoionization equilibrium.  The absorption lines are similar
to lower ionization absorption lines observed in the UV, although these UV lines
originate from a different location or phase of the absorbing medium.
Redshifted with respect to the absorption lines, emission from the \ion{O}{viii}
Ly$\alpha$ line as well as the \ion{O}{vii} triplet is visible.  The flux of
these lines is consistent with emission from the absorbing medium.  The
\ion{O}{vii} triplet intensity ratios demonstrate that photoionization dominates
and yield an upper limit to the electron density of $7\times 10^{16}$~m$^{-3}$.

\end{abstract}

\section{Introduction}

Low to medium energy resolution X-ray spectra of AGN such as obtained by the
Rosat or ASCA observatories showed the presence of warm absorbing material (see
references in Kaastra \cite{kaastra}).  This was deduced from broad band fits to
the continuum, showing a flux deficit at wavelengths shorter than the expected
edges of ions such as \ion{O}{vii} and \ion{O}{viii}.  The relation of this warm
X-ray absorber to the medium that produces narrow UV absorption lines in
\ion{C}{iv} or \ion{N}{v} is not clear, mainly due to a lack of sufficient
constraints in the X-ray band.
A major drawback of all previous X-ray studies of AGN has been the low spectral
resolution, making it hard to disentangle any emission line features from the
surrounding absorption edges, and prohibiting the measurements of Doppler shifts
or broadening.  With the Chandra spectrometers it is now possible for the first
time to obtain high-resolution X-ray spectra of AGN.

\section{Observations}

The present Chandra observations were obtained on December 11/12, 1999, with an
effective exposure time of 86400~s.  The detector used was the High Resolution
Camera (HRC-S) in combination with the Low Energy Transmission Grating (LETG).
The spectral resolution of the instrument is about 0.06~\AA\ and almost constant
over the entire wavelength range (1.5--180~\AA).  Event selection and background
subtraction were done using the same standard processing as used for the
first-light observation of Capella (Brinkman et al.  \cite{brinkman}).  The
wavelength scale is currently known to be accurate to within 15~m\AA.  The
efficiency calibration has not yet been finished, and our efficiency estimates
are based upon preflight estimates for wavelengths below 60~\AA\ and on inflight
calibration based upon data from Sirius~B for longer wavelengths.  We estimate
that the current effective area is accurate to about 20--30~\%, that it may show
large scale systematic variations within those limits, but it does not show
significant small scale variations.

The observed count spectrum was corrected for higher spectral order
contamination by subtracting at longer wavelengths the properly scaled observed
count spectrum at shorter wavelengths.  The spectrum was then converted to flux
units by dividing it by the effective area, and by correcting for the galactic
absorption of $1.65\times 10^{24}$~m$^{-2}$ (Nandra et al.  \cite{nandra}), as
well as for the cosmological redshift, for which we took the value of 0.01676
(Crenshaw et al.  \cite{crenshaw}).
\begin{figure}
\resizebox{\hsize}{!}{\includegraphics[angle=-90]{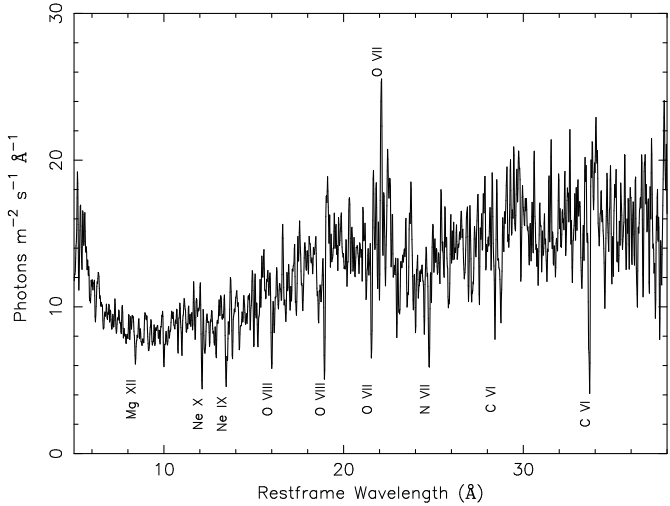}}
\caption{Chandra LETGS X-ray spectrum of NGC~5548,
corrected for order contamination, redshift and galactic
absorption.}
\label{fig:fig1}
\end{figure}
The spectrum in the 5--38~\AA\ range is shown in Fig.~\ref{fig:fig1}.  The
continuum is rather smooth; the reality of the large-scale structures cannot be
assessed completely at this moment given our current understanding of the
efficiency calibration of the instrument.  Nevertheless, there is no indication
for the presence of strong \ion{O}{vii} or \ion{O}{viii} K-shell absorption
edges at 16.77 and 14.23~\AA, respectively.  A more detailed discussion of the
spectrum, including the long-wavelength part will be given in a forthcoming
paper, when the full efficiency calibration of the instrument is available.

\subsection{Absorption lines}

The most striking feature of the spectrum is the presence of narrow absorption
lines, including the Lyman $\alpha$ and $\beta$ transitions of H-like C, N, O,
Ne and Mg, as well as the 2--1 resonance absorption line of He-like O and Ne.
We have searched the wavelength range of Fig.~\ref{fig:fig1} systematically for
absorption and emission lines, and found the lines listed in
Table~\ref{tab:data}.  In addition we provide data for some weaker features for
which the equivalent widths (or their upper limits) help to constrain models.
We give expected wavelength $\lambda_0$, the measured wavelength difference
$\Delta\lambda\equiv \lambda_0-\lambda_{\mathrm obs}/(1+z)$ (thereby accounting
for the cosmological redshift $z$), the equivalent line width $W$ (determined
from a gaussian fit to the line profile), and the proposed line identification.
A negative sign before an equivalent width indicates an emission line.
\begin{table}[!h]
\caption{Absorption and emission lines identified in NGC~5548.
Lines possibly blended by lines from other ions are indicated
by an asterisk.}
\label{tab:data}
\centerline{
\begin{tabular}{|rrrl|}
\hline
$\lambda_0$ (\AA)& $\Delta\lambda$ (m\AA) & W (m\AA)& Identification \\
\hline 
 5.217 &  57$\pm$ 25 &  19$\pm$ 15 & Si XIV 1s - 3p (Ly$\beta$)   \\
 6.182 &  12$\pm$ 16 &  18$\pm$  7 & Si XIV 1s - 2p (Ly$\alpha$)  \\
 8.421 &   1$\pm$ 11 &  24$\pm$  8 & Mg XII 1s - 2p (Ly$\alpha$)  \\
 9.169 &  -9$\pm$ 29 &  14$\pm$  9 & Mg XI 1s$^2$ - 1s2p $^1$P$_1$ (r)\\
 9.314 & -19$\pm$  0 &  -9$\pm$ 11 & Mg XI 1s$^2$ - 1s2s $^3$S$_1$ (f)\\
10.025 & -16$\pm$ 15 &  21$\pm$ 10 & Na XI 1s - 2p (Ly$\alpha$)    \\
11.003 &   8$\pm$ 13 &  21$\pm$ 10 & Na X 1s$^2$ - 1s2p $^1$P$_1$ (r) \\
12.134 &   0$\pm$  7 &  38$\pm$  9 & Ne X 1s - 2p (Ly$\alpha$) *   \\
12.274 &  13$\pm$ 21 &  26$\pm$  9 & Fe XVII 2p-4d *     \\
12.292 &  -5$\pm$ 21 &  26$\pm$  9 & Fe XXI 2p-3d *      \\
12.844 & -27$\pm$ 23 &  16$\pm$ 11 & Fe XX 2p-3d blend   \\
12.904 &  -4$\pm$ 12 &  25$\pm$ 11 & Fe XX 2p-3d         \\
13.447 &  21$\pm$  8 &  46$\pm$  9 & Ne IX 1s$^2$ - 1s2p $^1$P$_1$ (r) * \\
13.522 &  26$\pm$ 12 &  25$\pm$ 11 & Fe XIX 2p-3d blend  \\
13.698 &  24$\pm$ 32 & -18$\pm$ 13 & Ne IX 1s$^2$ - 1s2s $^3$S$_1$ (f) \\
13.826 &  59$\pm$ 11 &  26$\pm$ 10 & Fe XVII 2p-3p       \\
14.207 &   0$\pm$ 14 &  22$\pm$ 10 & Fe XVIII 2p-3d blend\\
15.014 &   5$\pm$ 12 &  19$\pm$ 10 & Fe XVII 2p-3d       \\
15.176 &  64$\pm$ 13 &  25$\pm$  9 & O VIII 1s - 4p (Ly$\gamma$)   \\
15.265 & -39$\pm$ 13 &  25$\pm$  9 & Fe XVII 2p-3d       \\
16.006 &   7$\pm$ 10 &  40$\pm$  9 & O VIII 1s - 3p (Ly$\beta$)     \\
16.612 &  10$\pm$ 11 & -45$\pm$ 15 & No id               \\
17.396 & -19$\pm$ 28 &  20$\pm$ 10 & O VII 1s$^2$ - 1s5p $^1$P$_1$ \\
17.768 & -25$\pm$ 17 &  22$\pm$ 10 & O VII 1s$^2$ - 1s4p $^1$P$_1$ \\
18.627 &  -7$\pm$ 17 &  21$\pm$ 10 & O VII 1s$^2$ - 1s3p $^1$P$_1$ \\
18.969 & -22$\pm$  7 &  55$\pm$ 10 & O VIII 1s - 2p (Ly$\alpha$)    \\
21.602 & -40$\pm$ 10 &  53$\pm$ 12 & O VII 1s$^2$ - 1s2p $^1$P$_1$ (r) \\
21.602 &  70$\pm$ 15 & -33$\pm$ 16 & O VII 1s$^2$ - 1s2p $^1$P$_1$ (r) \\
21.804 &  25$\pm$ 17 & -33$\pm$ 19 & O VII 1s$^2$ - 1s2p $^3$P$_{1,2}$(i) \\
22.101 &  11$\pm$ 11 & -64$\pm$ 19 & O VII 1s$^2$ - 1s2s $^3$S$_1$ (f) \\
24.781 &  -4$\pm$ 12 &  54$\pm$ 14 & N VII 1s - 2p (Ly$\alpha$)     \\
28.466 & -25$\pm$ 11 &  36$\pm$ 12 & C VI 1s - 3p (Ly$\beta$)      \\
33.736 & -36$\pm$ 10 &  99$\pm$ 17 & C VI 1s - 2p (Ly$\alpha$)       \\
38.950 & -88$\pm$ 25 &  35$\pm$ 22 & Fe XV 3s$^2$-3s5p      \\
39.146 &  25$\pm$ 25 & 164$\pm$ 61 & No id               \\
40.268 & -46$\pm$ 45 & 216$\pm$ 82 & C V 1s$^2$ - 1s2p $^1$P$_1$ (r) \\
50.350 & -11$\pm$ 32 &  47$\pm$ 30 & Fe XVI 3s-4p        \\
52.911 & -36$\pm$ 12 &  73$\pm$ 21 & Fe XV 3s$^2$-3s4p   \\
58.963 &  -9$\pm$ 52 &  48$\pm$ 40 & Fe XIV 3p-4d        \\
\hline\noalign{\smallskip}
\end{tabular}
}
\end{table}
The presence of these aborption lines can be seen as evidence for a warm,
absorbing medium in NGC~5548 along the line of sight towards the nucleus.  The
absorption can be very strong:  the core of the \ion{C}{vi} Ly$\alpha$ line for
example absorbs some 90~\% of the continuum, and this is just a lower limit,
since the true line profile is smeared out by the instrument.  That the optical
depth in some lines is considerable is evidenced by two facts:  firstly, the
equivalent width ratio of the Ly$\beta$ to Ly$\alpha$ lines of \ion{C}{vi} and
\ion{O}{viii} is much larger than the ratio of their oscillator strengths (0.079
to 0.417).  Secondly, we see absorption features of sodium, despite the fact
that the sodium abundance is 20 times smaller than, e.g., the magnesium
abundance.  All this can be explained if the line cores of the more abundant
elements are strongly saturated.

\subsection{Column densities}

Using the observed equivalent width $W$ of the absorption lines, it is possible
to derive the absorbing column density, assuming a gaussian velocity
distribution (standard deviation $\sigma_{\rm v}$) of the absorbing ions and
neglecting the scattered line emission contribution:
\begin{equation}
\label{eqn:w}
W = {\lambda \sigma_{\rm v}\over c} \int\limits_{-\infty}^{\infty}
[1-\exp (-\tau_0\mathrm{e}^{\displaystyle -y^2/2})]
{\mathrm d}y,
\end{equation}
with $\tau_0$ the optical depth of the line at the line center,
given by
\begin{equation}
\label{eqn:tau}
\tau_0 = 0.106 f N_{20} \lambda / \sigma_{\rm v,100}.
\end{equation}
Here $f$ is the oscillator strength, $\lambda$ the wavelength in \AA,
$\sigma_{\rm v,100}$ the velocity dispersion in units of 100~km/s and $N_{20}$ the
column density of the ion in units of $10^{20}$~m$^{-2}$.  Given a value for
$\sigma_{\rm v}$ and the measured equivalent width, these equations yield the
column density.  For some ions we have more than one absorption line identified,
and this allows us to constrain $\sigma_{\rm v}$.  From the \ion{O}{vii},
\ion{O}{viii} and \ion{C}{vi} ions we obtain $\sigma_{\rm v}$=140$\pm$30~km/s.
Using this value, we derive the column densities of Table~\ref{tab:col}.
\begin{table}[!h]
\caption{Derived column densities}
\label{tab:col}
\centerline{
\begin{tabular}{|lrr|lrr|}
\hline
ion & log $N$     & log $N$     & ion & log $N$   & log $N$        \\
    & observed    & model       &     & observed  & model   \\
    &  (m$^{-2}$) &  (m$^{-2}$) &     & (m$^{-2}$)&   (m$^{-2}$) \\
\hline
\ion{C}{vi}    & 21.2$\pm$0.3 & 21.2 & \ion{Si}{xiv}  & 22.6$\pm$1.6 & 19.8 \\
\ion{N}{vii}   & 21.1$\pm$0.5 & 21.0 & \ion{Fe}{xiv}  & 19.9$\pm$0.7 & 20.1 \\
\ion{O}{vii}   & 21.2$\pm$0.3 & 21.3 & \ion{Fe}{xv}   & 20.3$\pm$0.2 & 20.1 \\
\ion{O}{viii}  & 22.1$\pm$0.3 & 22.1 & \ion{Fe}{xvi}  & 20.4$\pm$0.5 & 19.6 \\
\ion{Ne}{ix}   & $<$22.8$\pm$1.2 & 21.1 & \ion{Fe}{xvii} & 20.5$\pm$0.3 & 20.5 \\
\ion{Ne}{x}    & $<$22.6$\pm$1.1 & 21.2 & \ion{Fe}{xviii}& 20.3$\pm$0.5 & 20.4 \\
\ion{Na}{x}    & 20.9$\pm$0.8 &  -   & \ion{Fe}{xix}  & 20.6$\pm$0.7 & 20.0 \\
\ion{Na}{xi}   & 21.4$\pm$1.0 &  -   & \ion{Fe}{xx}   & 20.9$\pm$0.5 & 19.3 \\
\ion{Mg}{xi}   & 20.8$\pm$0.7 & 20.8 & \ion{Fe}{xxi}  &$<$20.9$\pm$0.7 & 18.2 \\
\ion{Mg}{xii}  & 22.3$\pm$1.3 & 20.4 & & & \\
\hline\noalign{\smallskip}
\end{tabular}
}
\end{table}
We give the column density $N$ in logarithmic units.  The reason is the
relatively large inferred optical depth of some lines, e.g.  70 for the
\ion{O}{viii} Ly$\alpha$ line.  This makes the line core saturated and hence
significant changes in the column density lead to minor changes in the
equivalent width.

One of the most striking features of the spectrum is the absence of the oxygen
continuum absorption edges that were deduced from low resolution X-ray spectra
such as those acquired with Rosat (Nandra et al.  \cite{nandra}), ASCA (Fabian
et al.  \cite{fabian}) or BeppoSAX (Nicastro et al.  \cite{nicastro}).  The
absence is, however, consistent with the column densities derived above from the
line absorption.  We predict a jump of 11~\% at the \ion{O}{viii} edge
(14.23~\AA) and 4~\% at the \ion{O}{vii} edge (16.77~\AA), all within a factor
of 2.  We can measure any jumps near the edges with an accuracy of about 10~\%
of the continuum, but we find no evidence for an absorption edge; the data even
suggest a small emission edge (radiative recombination continuum) of
10$\pm$10~\%.

The column densities of the other ions for which we have absorption measurements
do not lead to significant absorption edges, except for \ion{Ne}{ix} and
\ion{Ne}{x}, which should be at the low side of their column density range in
order to avoid significant continuum absorption.  Note that the \ion{Ne}{x}
Ly$\alpha$ line has some blending from \ion{Fe}{xvii} 4d-2p; taking that into
account leads to a somewhat smaller column density.

We have made a set of runs using the XSTAR photoionization code (Kallman \&
Krolik \cite{kallman}), using solar abundances and the spectral shape as given
by Mathur et al.  (\cite{mathur}), normalised to 13
photons~m$^{-2}$s$^{-1}$\AA$^{-1}$ at 20~\AA.  We obtained a good overall
agreement with our measured column densities using a hydrogen column density of
$3\times 10^{25}$~m$^{-2}$ and $\xi=100\pm 25$ (in units of 10$^{-9}$~W\,m).
This column density is comparable to the value derived from earlier ASCA
observations (Fabian et al.  \cite{fabian}).  However, our ionization parameter
is significantly larger, having most of the oxygen as \ion{O}{viii} or
\ion{O}{ix}.  The plasma temperature of the absorber implied by the XSTAR model
is $2\times 10^{5}$~K.  The low temperatures imply that thermal contributions to
line broadening ($\sigma_{\rm v}$) are negligible.

\subsection{Velocity fields}

\begin{figure}
\resizebox{\hsize}{!}{\includegraphics[angle=-90]{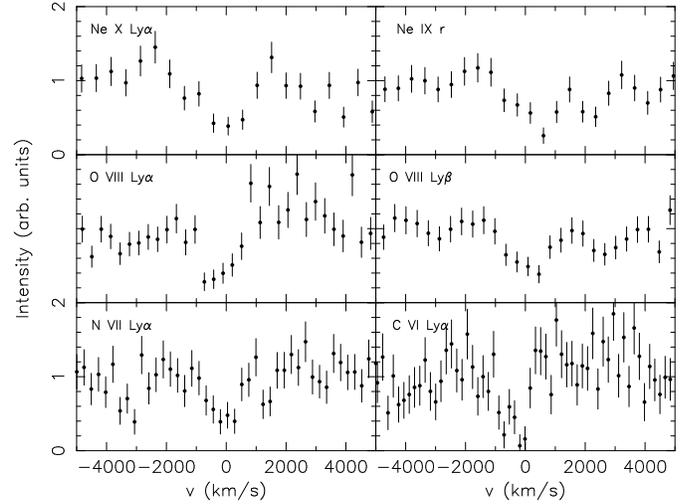}}
\caption{Line profile for six absorption lines. The intensity is scaled
to 1 in the $\pm$(5000--10000)~km/s range. Bin size is 0.02~\AA.}
\label{fig:fig2}
\end{figure}

The absorption lines appear to be blueshifted:  the average blueshift of the C,
N and O lines is 280$\pm$70~km/s.  There is some evidence that the lines are
broadened in proportion to their wavelengths, indicative of Doppler broadening.
Subtracting the instrumental line width ($\sigma$=0.023~\AA) yields for the
intrinsic line broadening a width ($\sigma$) of 270$\pm$100~km/s, somewhat
larger than the width of 140~km/s derived from line ratios (previous section).
This could indicate that the absorber consists of a few narrow components
($\sigma\sim$140~km/s), with different mean velocities and $\sigma\sim$270~km/s
for the ensemble.  As an illustration we show the velocity profile of six of the
strongest absorption lines in Fig.~\ref{fig:fig2}.  On the blue side, the line
profiles extend out to about 2000~km/s.  There is no clear evidence for the
presence of an underlying broad emission component for these lines, although for
\ion{O}{viii} and \ion{C}{vi} there appears to be an excess at the red side of
the absorption line.

\subsection{Emission lines}

\begin{figure}
\resizebox{\hsize}{!}{\includegraphics[angle=-90]{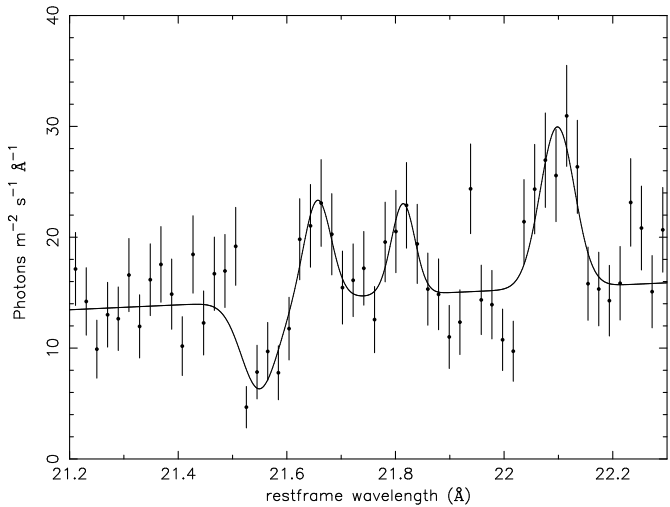}}
\caption{Spectrum near the \ion{O}{vii} triplet (data with error bars) plus
a simple model fit (continuum plus gaussian lines) for the main absorption
and emission components discussed in the text.}
\label{fig:fig3}
\end{figure}

The LETGS spectrum of NGC~5548 shows only a few emission lines.  Except for a
clear detection of the He-like triplet of \ion{O}{vii}, we only identified the
forbidden lines of the same triplets of \ion{Ne}{ix} and \ion{Mg}{xi}; these are
marginally detected.  Here we focus upon the \ion{O}{vii} triplet
(Fig.~\ref{fig:fig3}).  The forbidden ($f$) and intercombination ($i$) line are
not blue-shifted like the absorption lines but have a marginally significant
redshift of 200$\pm$130~km/s.  The ratio $i/f$ can be used as a density
diagnostic if the coupling between the upper levels ($2^3$S and $2^3$P) of $f$
and $i$ is determined by electron collisions and not by the external radiation
field from the central source.  For a photon flux of
50~photons\,m$^{-2}$s$^{-1}$\AA$^{-1}$ at 1600 \AA\ and the atomic parameters
taken from Porquet \& Dubau (\cite{porquet}) we estimate that this is the case
as long as $n_{\mathrm e} > 4\times 10^{14}$~m$^{-3}$.  The ratio $i/f$ does
depend only weakly upon the type of ionization balance:  Collisional ionization
equilibrium (CIE) or photoionization equilibrium (PIE) (Mewe \cite{mewe} and
Porquet \& Dubau \cite{porquet}).  From the observed value $i/f$ of
0.45$\pm$0.29 we derive an upper limit to the electron density $n_{\mathrm e}$
of $7\times 10^{16}$~m$^{-3}$.  The observed ratio $G=(i+f)/r$ is 3.2$\pm$1.5,
although this value might be somewhat smaller due to overlap of the $r$
absorption component.  For CIE plasmas, $G$ should be of order 1, while for PIE
plasmas, values larger than about 4 can be expected (Liedahl \cite{liedahl},
Porquet \& Dubau \cite{porquet}).  Thus, the \ion{O}{vii} triplet probably
originates from a photoionized plasma.
Does the emission from the triplet arise from the same medium that absorbs the
continuum?  Assuming that the absorber has the shape of a thin, spherical shell,
we calculate on the basis of recombination and the absorber parameters derived
in section 2.2 emission line intensities that agree within the error bars with
the measured intensities.  The upper limit for $n_{\mathrm e}$ found from the
$i/f$ ratio then implies a lower limit for the thickness of the shell of
$5\times 10^{8}$~m, and a distance from the central source of at least $8\times
10^{13}$~m.  Thus, both the absorption and emission of the \ion{O}{vii}
resonance line, as well as the emission from $i$ and $f$ may originate in the
same expanding shell.

\section{Discussion}

For the first time we see narrow absorption lines in the X-ray spectrum of an
AGN.  However narrow absorption lines in Seyfert galaxies have been seen before
in the UV band.  Shull and Sachs (\cite{shull}) discovered narrow absorption
features in the \ion{C}{iv} and \ion{N}{v} lines.  This was confirmed by Mathur
et al.  (\cite{mathur}) and studied in more detail by Mathur et al.
(\cite{mathur99}) and Crenshaw et al.  (\cite{crenshaw}).  These last authors
find at least 5 narrow absorption components in the \ion{C}{iv}~1550~\AA\, and
\ion{N}{v}~1240~\AA\, lines.  These components are broadened by
$\sigma_v$=20--80~km/s, somewhat smaller than we find.  The rms width of the
ensemble of UV absorption lines is 160~km/s for \ion{C}{iv} and 260~km/s for
\ion{N}{v}, consistent with the effective line width of 270$\pm$100~km/s that we
find from our Chandra data.  Also, the average blueshift of the UV absorption
lines ($-390$~km/s for \ion{C}{iv} and $-490$~km/s for \ion{N}{v}) is only
slightly larger than what we find for the C, N and O lines ($-280\pm 70$~km/s).
Note that our wavelength scale has residual uncertainties of 100--300~km/s for
most of our lines.

However, the column density of the lithium-like ions \ion{C}{iv} and \ion{N}{v}
as derived by Crenshaw et al.  is 100 times smaller than the column density of
the corresponding hydrogen-like ions that we find.  The difference may be
attributed to either time variability (low column density during the UV
observations), a high degree of ionization (hydrogenic ions dominating) or a
stratified absorber (with UV and X-ray absorption lines originating from
different zones).  We favour this last possibility.  This is supported by the
fact that our simulations with XSTAR imply \ion{C}{iv} and \ion{N}{v} columns
that are 100 times smaller than the measured values by Crenshaw et al.

Crenshaw \& Kraemer (\cite{crenshawk}) find that the weakest of the five
dynamical components (their number 1) in the UV absorption lines has the highest
outflow velocity ($-1056$~km/s).  Based upon the \ion{N}{v} to \ion{C}{iv}
ratio, they argue that this component has the highest ionization parameter and
could produce the oxygen continuum absorption edges as implied by the ASCA data.
Our modelling with XSTAR also predicts column densities of \ion{N}{v} and
\ion{C}{iv} close to the measured values for component 1.  But the outflow
velocity of the X-ray absorber that we find is significantly smaller than the
velocity of UV component 1.  However, Mathur et al.  (\cite{mathur99}) identify
component 3 ($-540$~km/s) as the most likely counterpart to the X-ray warm
absorber.  We conclude that the detailed relation between UV and X-ray absorbers
is still an open issue.

\begin{acknowledgements}
The Laboratory for Space Research Utrecht is supported
financially by NWO, the Netherlands Organization for Scientific
Research. Work at LLNL was performed under the auspices of the
U.S. Department of Energy, Contract No. W-7405-Eng-48.

\end{acknowledgements}

\end{document}